\documentclass[11pt, oneside]{article}   	
\usepackage{geometry}                		
\usepackage{graphicx}				
					          
\usepackage{amssymb}

\title{A quantum theory of distance along a curve}
\author{Ronald J. Adler* \\
Hansen Experimental Physics Laboratory, Gravity Probe B Mission, \\
and Kavli Institute for Particle Astrophysics and Cosmology,\\ Stanford University, Stanford, California 94309 and
\\Department of Physics and Astronomy, \\ San Francisco State University, San Francisco, California 94132}
\date{April 16, 2014}				
\begin{document}
\maketitle
\begin{abstract}
We present a quantum theory of distances along a curve, based on a linear line element that is equal to the operator square root of the quadratic metric of Riemannian geometry. Since the linear line element is an operator, we treat it according to the rules of quantum mechanics and interpret its eigenvalues as physically observable distances; the distance eigenvalues are naturally quantized. There are both positive and negative eigenvalues, which requires some interpretation. Multi-element curves are defined as direct sums of line elements, and behave much like systems of spin one half particles in a magnetic field. For a curve of many elements an entropy and energy and temperature are quite naturally defined, leading via standard statistical thermodynamics to a relation between the most probable curve length and temperature. That relation may be viewed as a universal heat-shrinking property of curves. At this stage of the theory we do not include bodies or particles in the mix, do not suggest field equations for the quantum geometry, and questions of interpretation remain. The theory might  conceivably be testable using observations of the early Universe, when the temperature of space was presumably quite high. In particular cosmogenesis may be thought of as time stopping at an infinite temperature of space as we go backwards in time to the beginning.  
\end{abstract}
Keywords General relativity $\cdot$ Non commutative geometry $\cdot$ Cosmology \\
*electronic mail address: adler@relgyro.stanford.edu and gyroron@gmail.com

\section{Introduction}

In Riemannian geometry, as used in classical general relativity (GR), the points of space are labelled by coordinates $x^\mu$ in an arbitrary way, for example Cartesian or spherical coordinates in three dimensional Euclidean 
space and Lorentz coordinates in four dimensional spacetime \cite{SW, ABS, MTW}. The coordinates serve only to label the points and have no intrinsic geometric or physical meaning. Because of this limited role of the coordinates it is obvious that any physical theory should be independent of the coordinate choice, or covariant. To relate the coordinate labels to distances we introduce a metric, by means of which the distance between two nearby points is given as a quadratic function of the coordinate differentials between the points, $ds^2=g_{\mu\nu} dx^\mu dx^\nu$. Thus only the combination of coordinates plus metric gives a meaning to distance. Of course only the local geometry is completely characterized in this way and the space can have various global topologies.

In this paper we will take the term ``metric" to mean both the matrix $g_{\mu \nu}$ of real numbers, and the functional form of the line element $ds^2=g_{\mu\nu}dx^\mu dx^\nu$, so the terms line element and metric will be used interchangeably. 

More general kinds of metric, such as used in a Finsler space, have been studied in differential geometry, but the choice of quadratic metric has dominated physics research. This has its basis in the Pythagorean theorem, and can be strongly motivated by physical considerations.\cite{ABS}  

In the present work we depart from using the quadratic metric by taking its square root to obtain a linear metric or line element, which is one in which the distance between two nearby points is a linear function of the coordinate differentials.\cite{Pagels} The tool we use to take the square root in three-dimensions is the Pauli algebra and in four-dimensions it is the Dirac algebra; both are examples of a Clifford algebra.\cite{Shankar, BjD} The price paid for taking the square root is that the linear line element is a matrix rather than a real number; we take this as strong motivation to interpret the line element as a Hermitian quantum operator and apply the rules of quantum mechanics (QM) to its analysis.\cite{Shankar} Exactly the same viewpoint and procedure allows one to obtain the Pauli equation for an electron in an electromagnetic field from the Schr\"{o}dinger equation for a free particle (Appendix A) and in a very similar way the Dirac equation for an electron in an electromagnetic field from the Klein-Gordon equation for a free particle (Appendix B).\cite{BjD}

With the line element viewed as a QM operator we are naturally led to study its eigenvalues and Hilbert space eigenvectors, and are faced with the questions of interpretation inherent in QM - but more subtle since distances at a very small scale, presumably Planckian, are involved and the meaning of an ``observation" is problematic.\cite{Shankar, Adler, Adler2} Almost needless to say the geometry is non-commutative since it involves a metric operator.\cite{Hogan, Hogan2, Connes}

There are three aspects of this paper which require assumptions, each assumption being somewhat more speculative than the preceding. The first is that we begin with a linear line element QM operator as noted in the above paragraphs. The second is that we assume invariant distances are multiples of some small fundamental distance - that is, distances are quantized.\cite{Oriti} There are well-known and strong motivations for the assumption that physical space is not well described by the mathematical continuum $\mathfrak{c}$ of real numbers at the fundamental scale.\cite{Adler, Adler2} Although there is no experimental evidence that physical space is not continuous there are various conceptual and theoretical problems associated with the continuum model. Many of the problems are related to the infinite density of degrees of freedom inherent in the continuum.\cite{Oriti2} For example the self-energy of a point charge in classical electromagnetism is infinite, and the analogous divergences of QFTs are only handled by clever artifices e.g. renormalization theory. Of course if no divergences occurred perturbative QFTs would still require renormalization, but the divergences make the process more delicate and difficult to reconcile with relativistic covariance. 
\cite{BjD, Peskin}

At a fundamental conceptual level there is an obvious intuitive strain associated with the real number continuum: the points on a line of any length can be put in one to one correspondence with the points of any other line, so any two lines have, by definition, the same number of points $\mathfrak{c}$. (As a trivial example we may place the points $x$ in the interval $[0,1]$ in one-to-one correspondence with the points $y$ in $[0,2]$ simply by taking $2x=y$.) Thus a line of Planck scale, $10^{-35} m$, has the same number of points as a line of Hubble scale, $10^{26} m$, so the number of degrees of freedom is the same for the smallest and largest distances that typically occur in present day physics. Another way to describe this property is that space is self-similar over an infinite range of scales. Taken at face value this property of lines seems very na\"{i}ve and simplistic and must strain our intuitive idea of what physical space ``should" be like, and it motivates a viewpoint in which ÒpointsÓ are replaced as fundamental things by ÒelementsÓ of a different sort, that is with discreteness or granularity. This is the second assumption of this paper, that a line of finite proper length contains an integer number of elements. The name line element thereby takes on an appropriate literal meaning.

The third assumption involves the interpretation of the theory, something which is still not entirely settled for QM even after  nearly a century of study. For our theory an analog of the Copenhagen interpretation would, in principle, involve measurements or observations of distances at the Planck scale, which is highly problematic. Instead we appeal to statistical thermodynamic ideas applied to macroscopic curves with many elements $N$ to study systems on a scale much larger than the elemental scale $\epsilon$. This approach is analogous to that used in the kinetic theory of the nineteenth century. Since we will deal with large $N$ it is natural to invoke the concept of entropy for a curve and use a statistical analysis based on the micro-canonical ensemble, in which the entropy is some function of the total energy and number of elements in the curve $S=S(E,N)$. We thus need to assume an effective geometric energy $E_g$ for a curve. Accordingly, the third fundamental assumption we make is that the geometric energy is proportional to the total length of the curve, $E_g \propto s$.

With the statistical thermodynamic approach it is straight-forward to calculate the entropy of a curve in terms of the energy and number of elements, and from that an effective geometric temperature. In fact the curve system behaves like a set of spins in a magnetic field. The curve thus shares with the spin system some peculiar features, such as the possibility that it has a negative temperature. The concept of temperature is thus generalized far beyond the simple ideas of kinetic theory in which temperature is the average kinetic energy of molecules in motion; however the assignment of a temperature to a curve is not any more abstract than the assignment of a temperature to a black hole surface, a purely geometric object. 

Although we assume a reasonable energy and temperature for a curve we do not yet offer any method to put a curve into a state with a specified energy or temperature. 

It is also straight-forward to apply the canonical ensemble approach to a curve, in which $N$ and $T_g$ are specified and the Helmholtz free energy $A(N,T_g)$ forms the basis for studying thermodynamic properties. 

One of our more amusing results is that the length of a curve in spacetime, that is its proper time or "meter stick distance," is related to its temperature by a remarkably simple equation that implies that the length decreases with temperature, or heat-shrinks. 

The paper is organized as follows. In sec.2 we obtain the linear line element by taking the square root of the quadratic line element, first for Euclidean three-space using the Pauli algebra and then for four dimensional spacetime using the Dirac Algebra. The eigenvalues of the line element are obtained in sec.3, and we also motivate and discuss the use of a distance quantum. In sec.4 we show that a line element has an intrinsic minimum transverse uncertainty due to its non-commutative nature; this is in close analogy with similar properties of angular momentum. Macroscopic curves are defined as direct sums of line elements and their properties are discussion in sec.5. In sec.6 we discuss the problem of interpretation of the distance quantum theory and our suggestion for a statistical thermodynamic interpretation of a multi-element curve.  Sec.7 lays out the framework for generalizing the theory to a curved Riemannian space. In sec.8 we summarize and speculate a bit more. Appendices A and B illustrate the use of the operator square root in obtaining the Pauli and Dirac equations from the Schr\"{o}edinger and Klein-Gordon equations for a free particle. Appendix C is a discussion of the eigenvectors of the Pauli and Dirac matrices.

It is important to note what we do not do in this work. For now we only discuss distances along curves and do not study areas or volumes that require systems of curves, for example some version of a spin network. We do not yet consider dynamical equations for the metric, which would presumably involve the Dirac matrices. 

We note that our theory does not appear to have any features in common with superstring theory. It obviously shares features with non-commutative geometry theories, such as those discussed by Hogan in references \cite{Hogan, Hogan2}. Finally our approach does share the goals of loop quantum gravity (LQG) in quantizing properties of geometry, but is more modest and phenomenological; more important, it is at present limited to consideration of the distances along curves and does not deal with other geometric entities such as areas and volumes nor with dynamics, whereas LQG begins with and is based on dynamical quantities; indeed LQG focuses on the affine connections rather than the metric, while our analysis focuses on the metric. Thus we deal with a structure that is entirely geometric in distinction to LQG, phenomenological and less ambitious and simpler than LQG, and at present is limited to only the curve length, a most crucial geometric property. Even more fundamental, our considerations do not explicitly deal with gravity, but only with the geometry of spacetime per se, so that we do not even speak of gravity at this stage; of course we expect that our considerations may have some bearing on gravity in the future. 

Finally, it is clear that our treatment of curves with associated entropy and temperature has similarities to thermogravity, as studied by Jacobson and Verlinde and others. \cite{Jacobson, Verlinde} Thermogravity involves entropy and temperature associated with a two-dimensional surface according to holography and entropic force ideas, and leads to heuristic derivations of Newtonian gravity and GR. It is possible that our present considerations of curve lengths might lead in a similar direction, but such questions are beyond the scope of the present paper. 

\section{The linear line element}

\subsection{Three dimensional Euclidean space}

We first discuss flat Euclidean space with Cartesian coordinates. The generalization to a curved Riemann space with any coordinate system is quite straight-forward as discussed in sec.7. The line element is 
\begin{equation}
\label{1}
{{ds_{cl}}^2=g_{ij}{dx^i}{dx^j}}=dx^2+dy^2+dz^2, \; {g_{ij}=\delta _{ij}},
\end{equation}
where we have denoted the metric by $g_{ij}$ for later convenience. To obtain a linear metric we take the square root of this using the Pauli algebra, defined by 
\begin{equation}
\label{2}
{\sigma_i \sigma_j = \delta_{ij} I + i {\epsilon_{ij}}^k \sigma _k},
\end{equation}
or equivalently 
\begin{equation}
\label{3}
{\{\sigma_i. \sigma_j\}/2 = \delta_{ij} I }, \; {[\sigma_i. \sigma_j]/2 =i {\epsilon_{ij}}^k \sigma _k }.
\end{equation}
The Pauli algebra is one example of a Clifford algebra. Then the linear line element, which we define as
\begin{equation}
\label{4}
{ds=\sigma_i dx^i},
\end{equation}
can be viewed as the square root of the quadratic metric (\ref{1}) since
\begin{equation}
\label{5}
{ds^2=\sigma_i \sigma_j dx^i dx^j =(1/2)\{ \sigma_i, \sigma_j \}dx^i dx^j =g_{ij}dx^i dx^j I=ds_{cl} ^2 I},
\end{equation}
which is simply the quadratic line element (\ref{1}) written twice. 

As noted in the introduction the operation of taking the root in this way has a noble pedigree; it is analogous to the way in which the Pauli equation describing an electron in an electromagnetic field can be obtained from the Schr\"{o}edinger equation for a free particle, including the correct g factor. (See appendix A.) 

One widely used representation of the Pauli matrices, in which $\sigma_3$ is diagonal, is \cite{BjD}
\begin{equation}
\label{6}
\sigma_1= \left( \begin{array}{cc} 0 & 1 \\ 1 & 0  \end{array} \right), \;  \sigma_2= \left( \begin{array}{cc} 0 & -i \\ i & 0  \end{array} \right), \; \sigma_3= \left( \begin{array}{cc} 1 & 0 \\ 0 & -1  \end{array} \right). 
\end{equation} 
We write the coordinate differential in terms of a unit direction vector and magnitude as $\vec dx = \hat n \epsilon$;  then the line element in terms of Cartesian and spherical coordinates is
\begin{equation}
\label{7}
ds=\epsilon (\hat n \cdot \sigma )=\epsilon \left( \begin{array}{cc} n^3 & n^1-in^2 \\ n^1+in^2 & -n^3  \end{array} \right)=\epsilon \left( \begin{array}{cc} \cos\theta & \sin\theta e^{-i\phi} \\ \sin\theta e^{i\phi}  & -\cos\theta  \end{array} \right).
\end{equation} 
The use of spherical coordinates will later prove to be convenient. The matrix line element (\ref{7}) will be interpreted as a quantum operator and the use of spherical coordinates for the direction will be useful in later sections. 

Introduction of the parameter $\epsilon$ is at this point a mathematical artifice of convenience. We will later, in sec.3, discuss it further in a physics context as an alternative to the simplistic idea that space and spacetime are described by the mathematical continuum. When we consider curves composed of many line elements in sec.5 it will be seen as a near necessity in defining the curve length operator. 

\subsection{Four dimensional spacetime}

We next discuss the flat spacetime of special relativity, using Lorentz coordinates, in close analogy with the discussion of the preceding section. The generalization to curved Riemannian spacetime is discussed in sec.7. The line element is 
\begin{equation}
\label{8}
{{ds_{L}}^2=g_{\mu \nu}{dx^{\mu}}{dx^{\nu}}}, \; {g_{\mu \nu}=\eta _{\mu \nu}},
\end{equation}
where $\eta_{\mu \nu}$ is the Lorentz metric, and we use the signature $(+,-,-,-)$. To obtain a linear metric we take the square root of this using the Dirac algebra, defined by 
\begin{equation}
\label{9}
\{\gamma_\mu, \gamma_\nu\} = \eta_{\mu\nu} I , \; (i/2)[\gamma_\mu, \gamma_\nu] \equiv \sigma_{\mu \nu}.
\end{equation}
The Dirac algebra is another example of a Clifford algebra. The linear metric is accordingly defined as
\begin{equation}
\label{10}
{ds=\gamma_{\mu} dx^{\mu}},
\end{equation}
which can be viewed as the square root of the quadratic metric (8) since
\begin{equation}
\label{11}
{ds^2=\gamma_{\mu} \gamma_{\nu}dx^{\mu} dx^{\nu} =(1/2) \{\gamma_{\mu}, \gamma_{\nu} \}dx^{\mu} dx^{\nu}= g_{\mu\nu}dx^{\mu} dx^{nu} I=ds_{L} ^2 I},
\end{equation}
which is simply (\ref{8}) written four times. \cite{Pagels}

As in three dimensions the operation of taking the root in this way also has a noble pedigree; it is the way in which the Dirac equation describing an electron in an electromagnetic field can be obtained from the Klein-gordon equation for a free particle, including the correct g factor. (See appendix B.) One representation of the Dirac  matrices is particularly convenient for our analysis, 
\begin{equation}
\label{12}
\gamma_0= \left( \begin{array}{cc} 0 & I \\ I & 0  \end{array} \right), \; 
\gamma_j= \left( \begin{array}{cc} 0 & \sigma_j \\ -\sigma_j & 0  \end{array} \right), \;
\sigma_{0j}= \left( \begin{array}{cc} -i \sigma_j & 0\\ 0 & i\sigma_j   \end{array} \right), \;
 \sigma_{ij}=i \left( \begin{array}{cc} \sigma_k & 0 \\ 0 &\sigma_k \end{array} \right), 
\end{equation} 
where $I$ and $\sigma_j$ are $2 \times 2$ matrices and $i,j,k$ are cyclic. \cite{Feynman}

In analogy with three dimensions we denote the coordinate displacement in terms of a direction four-vector $n^\mu=(n^0,  n^j)$ and magnitude parameter $\epsilon$ as $dx^\mu = n^{\mu} \epsilon$. We also introduce the convention 
$\sigma_{\mu}=(I,\sigma_j)$ and $\tilde \sigma_{ \mu}=(I,-\sigma_j)$ and use the representation (\ref{12}) to write the linear element as
\begin{eqnarray*}
\label{13}
ds=\epsilon (n^\mu \cdot \gamma_{\mu} )=\epsilon \left( \begin{array}{cc} 0 &  n^{\mu}\sigma_{\mu} \\ n^{\mu} \tilde\sigma_{\mu} & 0 \end{array} \right), 
\end{eqnarray*}
\begin{eqnarray*}
\label{13}
n^{\mu} \sigma_{\mu} = n^0 I + \vec n \cdot \sigma, \; \;
n^{\mu} \tilde \sigma_{\mu} = n^0 I - \vec n \cdot \sigma,
\end{eqnarray*} 
\begin{equation}
\label{13}
(n^{\mu} \sigma_{\mu})(n^{\nu} \tilde \sigma_{\nu})= (n^{\omega}n_{\omega})I .
\end{equation} 
The $4 \times 4$ matrix line element in (\ref{13}) will be interpreted as a quantum operator; its expression in terms of 
$n^j  \sigma_j=\vec n \cdot \sigma$ will be very convenient. The normalization of $n^\mu$ will be discussed in the next section.

The procedure used in the last two sections can be applied to any Riemann space that admits a Clifford algebra and for which the linear line element can be defined, as we will discuss in sec.7. 

\section{Quantum properties of the line element operator}

\subsection{Eigenvalues in three dimensions}

In quantum theory the quantities that can be measured and compared with observation are the eigenvalues of Hermitian operators and various matrix elements of the operators. Thus we first study the eigenvalue problem for the operator $\sigma \cdot \hat n$ in (\ref{7}) for three dimensions, which is the same as the QM spin eigenvalue problem for spin in the direction $\hat n$. Since $(\sigma \cdot \hat n)^2= I$ and $Tr(\sigma \cdot \hat n)=0$ the eigenvalues must be $+1$ and $-1$. Thus the eigenvalues of the line element are 
\begin{equation}
\label{14}
\lambda=\epsilon, \; -\epsilon.
\end{equation} 

The negative eigenvalues need interpretation. Their meaning for the length of a curve composed of many line elements will be discussed in later sections. The eigenvectors are readily obtained from (\ref{7}) but will not play an important role in later development. They are discussed in Appendix C. 

\subsection{Eigenvalues in four dimensions}

The four-vector $n^\mu$ in $dx^\mu = n^{\mu} \epsilon$ will have a norm which we choose in the conventional way 
\begin{eqnarray*}
\label{15}
n^\mu n_{\mu}=\eta,
\end{eqnarray*} 
\begin{equation}
\label{15}
timelike: \eta=1, \; \; spacelike: \eta=-1, \; \; null: \eta=0.
\end{equation} 
A convenient way to write $n^\mu$ is 
\begin{equation}
\label{16}
n^{\mu}=(\sqrt{\eta + {\vec n}^2},\vec n ),
\end{equation} 
so we see how $n^\mu$ is determined entirely by $\eta$ and $\vec n$.

The eigenvalues $\lambda$ of the line element $\gamma_{\mu} n^{\mu} \epsilon$ in (\ref{13}) depend on $\eta$ and 
$\epsilon$, and the standard procedure gives the characteristic equation
\begin{eqnarray*}
\label{17}
(\lambda ^2 - \eta \epsilon^2)^2=0 \;
\end{eqnarray*} 
\begin{equation}
\label{17a}
timelike: \lambda=+\epsilon, +\epsilon, -\epsilon, -\epsilon \;  
spacelike: \lambda=+ i\epsilon, +i\epsilon, -i\epsilon, -i \epsilon, \:
null: \lambda=0, 0,0,0.
\end{equation}
In four dimensional spacetime with signature $(+,-,-,-)$ a timeline interval corresponds to a proper time $\sqrt ds^2$ while a spacelike interval corresponds to a ``meter stick distance" $\sqrt {-ds^2}$, so the geometric meaning of the eigenvalues $\pm \epsilon, \pm i 
\epsilon$ is clear. 

In summary, the eigenvalues of the spacetime line element $ds$ come in pairs, $\pm \epsilon$ for a timelike direction, $\pm i \epsilon$ for a spacelike direction, and $0$ for a null direction. That is the line element is a two level quantum system just as it is in three dimensions, and therefore behaves like non-relativistic spin; the difference is mainly that the four-dimensional system has two-fold degeneracy and null eigenvalues.   

\subsection{The quantum of distance}

Our assumption of a quantum of distance has already been discussed in the introduction. It is certainly reasonable to take a curve to be made up of elements of some small length $\epsilon$, presumably of order the Planck distance of about $10^{-35}m$, at which scale it is widely thought that the spacetime continuum should be re-examined. In particular the generalized uncertainty principle provides motivation for the assumption of some sort of granularity. \cite{Adler3} Thus we will view a curve as being made up of a sum of such elements rather than an integral. The analogy with Planck's assumption that black body radiation interacts only in quanta is clear. Moreover, associating a quantum with the line element satisfies the obvious desire for it to be an invariant. Thus we will henceforth take the relation $dx^{\mu}=\epsilon n^{\mu}$ to have physical meaning rather than being a mere artifice to specify the direction of the line element; the assumption carries over to the covariant theory as discussed in sec.7. When we discuss multi-segment curves in sec.5 it will be seen to be a near necessity to view a curve as made up of finite length line elements rather than an integral. 

It will become clear that in both three and four dimensions a curve will behave like a set of spins, a very large set for any macroscopic curve and indeed for any curve we might encounter in present day physics. 

\section{Transverse uncertainty of the line element}

\subsection{Transverse uncertainty in three dimensions}

In three dimensions the line element is a quantum operator, which implies an intrinsic uncertainty in the direction of the line $\hat n$. The uncertainty is due to the non-commutativity of the operators for the three orthogonal directions. 
We can see this in three different ways. 

First consider a curve $C$ going through a point of interest, and line up the $z$ or 3 axis with the curve. See Fig.1. The line element along C and the corresponding eigenvector in the positive direction are then 
\begin{equation}
\label{18}
ds_c=ds_3=\epsilon \sigma_3, \; |+,\hat z \rangle = \left( \begin{array}{c} 1 \\ 0 \end{array} \right).
\end{equation}
Similarly the line element transverse to C, say in the $x$ or 1 direction, has the operator
\begin{equation}
\label{19}
ds_1=\epsilon \sigma_1
\end{equation}
Then the variance (square of the standard deviation) in the 1 direction, $Var(ds_1)$, for the curve eigenvector state in (18), is easy to calculate since $\sigma^2 =I$; it is
\begin{equation}
\label{20}
Var(ds_1)=\langle (ds_1)^2 \rangle - \langle ds_1 \rangle ^2 = \epsilon ^2,
\end{equation}
and similarly for the $x$ or 2 direction. Thus the standard deviations are 
\begin{equation}
\label{21}
SD(ds_1)=SD(ds_2)=\epsilon.
\end{equation}
This we interpret as an uncertainty in the direction of the curve $C$, or as a transverse uncertainty or quantum width. The standard deviation along the curve is of course equal to zero since the vector (\ref{18}) is an eigenvector of the line element along $C$. 

For a second way to see the transverse uncertainty we observe that the commutator of line element operators in the 1 and 2 directions is not zero, 
\begin{equation}
\label{22}
[ds_1,ds_2]=2i \epsilon^2 \sigma_3.
\end{equation}
By the standard methods of QM we obtain from (\ref{22}) an uncertainty principle for the line elements in the 1 and 2 directions
\begin{equation}
\label{23}
SD(ds_1)SD(ds_2)\geq \langle [ds_1,ds_2]/2i \rangle =\epsilon^2 \langle \sigma_3 \rangle =\epsilon^2, 
\end{equation}
in agreement with our previous result (\ref{21}).

For a third view of the transverse uncertainty we invoke the heuristic vector model of angular momentum. The eigenvector of the line element $ds$ is an eigenstate of the 3 component $\epsilon \sigma_3$ with eigenvalue $\epsilon$ and also of the square $ds^2=\epsilon^2 ( {\sigma_1}^2+ {\sigma_2}^2+ {\sigma_3}^2)=3 \epsilon ^2I$. We therefore visualize the line element as having a 3 component $\epsilon$ and a length of $\sqrt 3 \epsilon$ so it has a component in the orthogonal direction of $\sqrt 2 \epsilon$. Thus we can visualize the vector as lying on a cone around the 3 axis with a base radius of $\sqrt 2 \epsilon$. (See Fig.1.)

All of the above arguments lead to the same result, that a line element has a transverse uncertainty of order $\epsilon$. We will do a similar analysis for a curve composed of more than one line element segment in sec.5.

\subsection{Transverse uncertainty in four dimensions}

Let us first consider the case of a line element in the 0 or time direction, corresponding to the path of a particle at rest. Then $n^\mu=(1,0,0,0)$, and 
\begin{equation}
\label{24}
ds_0=\epsilon \gamma_0
\end{equation}
It is easy to see from the $\gamma_0$ in (\ref{12}) that any vector of the form 
\begin{equation}
\label{25}
V= \left( \begin{array}{c} \alpha \\ \alpha \end{array} \right),
\end{equation}
is an eigenvector, where $\alpha$ is any normalizable two-spinor. Then the variance of the line element $\epsilon \gamma_j$ in any spatial direction $j$ is easy to calculate since 
\begin{equation}
\label{26}
ds_j^2=-\epsilon^2I.
\end{equation}
The negative sign is, of course, an artifact of our choice of signature as we noted previously, so the variance must be defined as an absolute quantity. We thus obtain
\begin{equation}
\label{27}
Var(ds_j)=\langle (ds_j)^2 \rangle - \langle ds_j \rangle ^2 = \epsilon ^2, \; SD(ds_j) = \epsilon 
\end{equation}
That is, the line corresponding to a stationary particle has an uncertainty in the spatial direction, transverse to the curve direction, and is spatially spherically symmetric. 

We next consider a spacelike line element lined up with the $z$ axis. Then 
\begin{equation}
\label{28}
ds_3=\epsilon \gamma_3,
\end{equation}
and the corresponding eigenvector is easily obtained from (12) and is 
\begin{equation}
\label{29}
V= \left( \begin{array}{c} |+, \hat z \rangle \\ i|+, \hat z \rangle \end{array} \right).
\end{equation}
The calculation of the standard deviation in the 0 or time direction and the orthogonal 1 and 2 space directions proceeds as above, with the result
\begin{equation}
\label{30}
Var(ds_0)=Var(ds_1)=Var(ds_2)=\epsilon ^2\; ,  \; SD(ds_0) = SD(ds_1)=SD(ds_2) = \epsilon. 
\end{equation}

Thus in summary the various line elements we have analyzed all have transverse quantum uncertainties equal to the distance quantum $\epsilon$ in both three and four dimension. 

\section{Multi-segment curves}

\subsection{A curve as a direct sum of line elements}

In classical physics the length of a curve is an integral of the root of the line element $\sqrt{ds^2}$ for a timelike curve
or $\sqrt{-ds^2}$ for a spacelike curve. To define the length operator for a curve we will naturally use the direct sum of line element operators; this is straight-forward if the curve is composed of discrete elements as discussed in sec.3.3. It would be cumbersome to use a continuous sum in the definition, especially with regard to the eigenvectors. We consider this a strong mathematical motivation for the use of a small finite $\epsilon$ parameter, but not an absolute necessity. 

Thus we define the length of a curve as the direct sum of line element operators over $N$ segments; the curve itself is defined in terms of the $N$ quantities $\vec n$ or $n^{\mu}$  respectively in three or four dimensions.  Since each line element is a two level system the curve behaves like a chain of spins, as shown in Fig.2. 

Explicitly we write the curve length operator as 
\begin{equation}
\label{31}
3D: s=\displaystyle\sum_{j=1}^N ds = \epsilon \displaystyle\sum_{j=1}^N (\hat n_j \cdot \sigma) 
= \epsilon(\hat n_1 \cdot \sigma \oplus \cdot \cdot \cdot \oplus \hat n_N \cdot \sigma) ,
\end{equation}
\begin{equation}
\label{32}
4D: s=\displaystyle\sum_{j=1}^N ds = \epsilon\displaystyle\sum_{j=1}^N  ( n^{\mu} \gamma_{\mu}) = \epsilon( {n_1}^{\mu} \gamma_{\mu} \oplus \cdot \cdot \cdot \oplus {n_N}^{\mu} \gamma_{\mu} ). 
\end{equation}
We can also think of the direct sum as providing a physical definition of the magnitude of the distance quantum; $\epsilon$ is the maximum distance that makes the line element operators in (\ref{31}) or (\ref{32}) independent. 

Since the line elements are all two level systems we see that the length operator is the same as the Hamiltonian for a string of spins in a magnetic field with constant magnitude but varying direction; the varying direction does not complicate the QM problem since each line element is independent.  

\subsection{Length eigenvalues and eigenvectors of a curve} 

Since the curve length operator $s$ is a direct sum its eigenvalues are sums of the individual line element eigenvalues, $\pm \epsilon$ for a timelike curve. Thus the eigenvalues of $s$ are 
\begin{equation}
\label{33}
l=\epsilon (N_+ - N_-)
\end{equation}
where $N_+$ is the number of positive eigenvalues and $N_-$ is the number of negative eigenvalues, and $N_+ + N_ - =N$. Thus the length eigenvalues run from $-N\epsilon$ to $N\epsilon$. Classical curves do not have a negative length so the appearance of negative values for $l$ requires interpretation, as we will discuss in sec.6. \\

The eigenvectors of a direct sum can be written as products of the individual eigenvectors. Thus the eigenvectors of $s$ in three dimensions are simply gotten from Appendix C for the three dimensional case, 
\begin{equation}
\label{34}
|\psi \rangle = \displaystyle \prod_{j=1}^N |\pm, \hat n_j \rangle = |\pm, \hat n_1 \rangle \otimes \cdot \cdot \cdot |\pm, \hat n_N \rangle ,
\end{equation}
and similarly for the four dimensional case. The vectors in (\ref{34}) form a convenient basis in which the length of the curve is sharp in the QM sense and has observable values given in (\ref{33}).

\subsection{Transverse uncertainty of a curve}

In sec.4 we found that a line element of a curve $C$ has a transverse quantum uncertainty;  that is, the line element perpendicular to the curve has a variance $\epsilon ^2$ and standard deviation $\epsilon$ for the cases we looked at in both three and four dimensions. From the way in which we defined a curve length as the direct sum of  independent line elements we can obtain a similar result for the multi-segment curve. 

The perpendicular width of a curve $C$ will behave like a sum of independent random variables. It is well-known that the variance of a sum of independent random variables is the sum of the individual variances, so we see that
\begin{equation}
\label{35}
Var(s_{\bot})=N\epsilon ^2
\end{equation}
For the maximum length curve, that is the maximum $l$ in (33), we have $L=N \epsilon$ and thus we can express the variance and standard deviations perpendicular to $C$ as
\begin{equation}
\label{36}
Var(s_{\bot})=(L/\epsilon) \epsilon ^2=L \epsilon , \; \; SD(s_{\bot})=\sqrt{L \epsilon}.
\end{equation} 
This is a rather obvious result of the independence of the individual line elements. 

From (\ref{36}) we see that the transverse uncertainty in the curve can be much greater than that for the line element, which might be physically interesting. The transverse uncertainty (\ref{36})  depends on the length $L$ of the timelike curve. If we are interested in physics and a particle such as an electron it is not obvious what value we should use for 
$L$ . One might argue that it should be c times the lifetime of the electron, that is the age of the Universe, in which case the uncertainty would be of order $10^{-5} m$, which is too large to be reasonable. It seems more reasonable that it should be the decoherence time for the electron's quantum state, which is very much shorter. This is a physics question that transcends the geometric study we are presently making, so for now we offer no definitive answer. 

\section {Physical interpretation and the statistical thermodynamic point of view}

\subsection{The basic question of interpretation}

The interpretation of non-relativistic QM remains a topic of debate nearly a century after the theory was formulated. The original Copenhagen interpretation provides a way to relate the theory to experimental and observational results via some basic assumptions about the Hamiltonian operators $\Lambda$ of the theory. For our present purposes the important assumptions, which are unlikely to change with improvements in the epistemology, are: (1) The only values that can be ``observed in the lab" are the eigenvalues $\lambda_j$ of the operator $\Lambda$; (2) When observing the eigenvalues $\lambda_j$ for a system in a state $| \psi \rangle$ the expectation value (or average $\lambda_j$) is 
$\langle \psi | \Lambda | \psi \rangle$. In the present context of a curve operator the application of these rules is quite problematic since we do not have an obvious way to measure a curve length with any sort of clock or meter stick, especially since the relevant length scale is, presumably, Planckian. We will not consider such small scale measurement in this work.

Instead of attacking the presently intractable problem of measuring Planckian distances we will adopt a point of view analogous to that of theorists developing the kinetic theory of gases in the nineteenth century: while remaining ignorant of the internal structure of gas molecules theorists calculated interesting properties of large numbers of molecules based only on their energies and statistical assumptions. It was a highly successful endeavor in establishing the existence of atoms and clarifying the nature of thermodynamics and the concept of temperature. 

As we have previously noted, a curve in either three or four dimensions behaves like a collections of spins, so what we wish to do is essentially to analyze the statistical thermodynamics of a collection of 2-level objects, such as spin half particles in a magnetic field. 
 
\subsection{Entropy, energy and length of a curve} 

The two-level nature of the line element is its most important property in either three or four dimensions, and in this section we will only consider timelike curves in four dimensional spacetime. According to (\ref{33}) the length of a curve with $N$ line element segments will be a maximum of $N \epsilon$ when all the line element eigenvalues are $+\epsilon$, which we may call spin up. If one segment has a reversed spin down eigenvalue $-\epsilon$ the length will be $\epsilon (N-2)$ and so forth, so that if $k$ segments are spin down the length will be $\epsilon (N-2k)$. Moreover the number of ways that the $k$ spin down segments can be chosen from the $N$ segments is 
\begin{equation}
\label{37}
\Omega =  \left( \begin{array}{c} N \\ k \end{array} \right)=\frac{N!}{k! (N-k)!}.
\end{equation} 
This is the degeneracy of the curve, the number of quantum states with the same length eigenvalue $l=(N-2k) \epsilon$. \cite{SM}

The smallest distances associated with present day physics are typically of order $10^{-17}m$ or some 18 orders of magnitude larger than what we expect for the distance quantum $\epsilon \approx 10^{-35} m$, so $N$ is typically very large. It is thus quite natural to associate an entropy with a curve according to Boltzmann's relation, with the degeneracy in (\ref{37}) being the number of micro-states corresponding to a single length macro-state, so
\begin{equation}
\label{38}
S = \ln(\Omega).
\end{equation} 
From (\ref{38}) and (\ref{37}) we may calculate the entropy for large values of $N$ and $k$ with the aid of Stirling's formula to be
\begin{equation}
\label{39}
S = N\ln (N)-k \ln (k)-(N-k)\ln(N-k).
\end{equation} 

To complete the definition of entropy as a function of the number of states and total energy (the micro-canonical ensemble context) we need to assign an effective energy to the curve. As the simplest choice we will take the effective energy of a curve to be proportional to its length; also, in order to make the maximum length eigenvalue correspond to the ground state of the curve we choose the proportionality constant to be negative. Thus we choose an effective Hamiltonian and associated eigenvalues to be 
\begin{equation}
\label{40}
H_{ef}=-(E_{f}/\epsilon)s=E_{f} \displaystyle\sum_{j=1}^N (\hat n_j \cdot \sigma) , \; \; E=-E_{f}(N-2k).
\end{equation} 
On dimensional grounds we might expect the fundamental energy $E_{f}$ to be of order the Planck energy $\approx 10^{19}GeV$; however this is not necessary, as we will discuss below. \cite{alphag}

From the expressions (\ref{39}) and (\ref{40}) we now have the entropy expressed as a function of $N$ and $E$. There is no volume associated with the curve, so $S=S(N, E)$. This allows us to define a temperature and chemical potential for the curve, via the standard thermodynamic relations 
\begin{equation}
\label{41}
\frac{1}{T} =\left(\frac{\partial S}{\partial E} \right )_N, \; \frac{\mu}{T}=-\left( \frac{\partial S}{\partial N}\right )_E, 
\end{equation}
and since there is no volume associated with the curve the pressure is zero. Then from (39) and (40) we obtain for the temperature
\begin{equation}
\label{42}
\frac{1}{T} =\left(\frac{\partial S}{\partial k} \right )_N \left(\frac{\partial k}{\partial E} \right )
=\frac{1}{2E_f}\ln \left( \frac{N-E/E_f}{N+E/E_f} \right).
\end{equation}

The behavior of the curve system as a function of $k$ is the same as that of a system of spins in a magnetic field, which has some interesting and peculiar properties. For $k=0$ the entropy is zero, the energy is the ground state, the length is maximum, and the temperature is zero, 
\begin{equation}
\label{43}
k=0: \;E_{gnd}=-E_f N\; , \; l_{gnd} =N\epsilon , \; T=0 .
\end{equation}
As $k$ increases the entropy increases, the energy increases, the length decreases, and the temperature rises. When half of the spins are down, that is $k=N/2$, the entropy reaches a maximum, the energy is zero, the length is zero, and the temperature becomes infinite, 
\begin{equation}
\label{44}
k=N/2: \;E=0 , \; l =0, \; T=\infty .
\end{equation}
As $k$ increases beyond $N/2$ the entropy decreases, the energy increases, the length become negative, and the temperature become negative. Finally at $k=N$, when all the spins are down, the entropy reaches zero, the energy is maximum, and the length is most negative,
\begin{equation}
\label{45}
k=N: \;E=E_f N , \; l =-N \epsilon, \; T=0.
\end{equation}

This odd behavior is familiar for the analogous system of spins, and the same behavior occurs for any finite system with two energy levels. Indeed the same qualitative behavior occurs for any finite number of levels for a finite number of elements: when enough energy is put into the system the energy levels become equally populated, corresponding to infinite temperature, and when yet more energy is put into the system the higher levels must become more populated, corresponding to negative temperature. The infinite temperature does not indicate any sort of physical singularity, only that all levels of the system are equally populated. 

The expression (\ref{42}) giving the temperature as a function of energy may be inverted to give an elegant relation for the length of the curve 
\begin{equation}
\label{46}
E=-E_f N \tanh(E_f /T) , \; l=N\epsilon \tanh (E_f /T). 
\end{equation}
The second relation in (\ref{46}) will be encountered again in the next section when we take an alternative approach to the statistical thermodynamics of the curve. 

Previously we noted that the negative length eigenvalues appear to require interpretation. Now we see that to populate the negative levels significantly requires a very high or infinite temperature. This provides at least a partial answer to the interpretation question, but it is likely to be too simplistic. 

In closing this section we summarize the qualitative behavior of the curve length in ({\ref 46}): for temperature very small compared to $E_f$ the length is simply the ``classical" or maximum value $L=N\epsilon$, and as the temperature rises above $E_f$ the length shrinks toward zero. That is, the behavior is roughly like that of ``heat-shrink" plastics. This behavior is curiously unlike that of many other physical systems, for which quantum behavior becomes important at low temperatures when the de Broglie wavelengths of the constituent particles overlap; black holes in the process of evaporation also share the curious behavior, that quantum effects become more important the higher the temperature. 

\subsection{Alternative approach to curve thermodynamics}

In the previous subsection we treated the thermodynamics of the curve using the entropy as a function of the energy and number of line elements of the curve, $S(E,N)$. Alternatively we may formulate the theory in terms of the temperature of the curve  and the number of line elements by using the Helmholtz free energy, $A(T,N)$, which is obtained by a Legendre transform of $S(E,N)$. In this formulation, based on the canonical ensemble, the relative probability that the curve has an energy $E$ is the usual Boltzmann factor times the degeneracy of the the macro-state or length of the curve in (\ref{37}); that is, the probability that $k$ of the spins are down is
\begin{equation}
\label{47}
W(k)= \left( \begin{array}{c} N \\ k \end{array} \right)e^{-E/T}= \left( \begin{array}{c} N \\ k \end{array} \right)e^{E_{f}(N-2k)/T}.
\end{equation}
This probability distribution is very sharply peaked at its maximum if $N$ is large, as illustrated in Fig.3 for a modest value $N=30$. The most likely $k_{m}$ is easily found from (\ref{47}) and Stirling's formula; $k_m$ and the corresponding most probable length $l_m$ are
\begin{equation}
\label{48}
k_{m}=N/(1+e^{2E_f /T}) , \; l_m=\epsilon N \tanh(E_f/T),
\end{equation}
which agrees with the previous result (\ref{46}). 

The sharpness of the peak at $k_m$, as shown in Fig.3,  can be estimated by a standard method; we fit a parabola to the curve at its maximum and calculate from it the full width at half maximum, to find
\begin{equation}
\label{49}
\Delta k=\sqrt{N}/[2 \cosh (E_f /T)] 
\end{equation}
For temperature small compared to the fundamental energy the peak is thus very narrow. 

For a numerical example of the above results let us take a macroscopic curve with $N \approx 10^{36}$ line elements, 
$\epsilon\approx 10^{-35}m$, a fundamental energy of Planck scale $E_f\approx 10^{19}GeV$, and a curve temperature of about the GUT scale of particle theory, $10^{17} GeV$, or about $1\%$ of the Planck energy. Then the width of the peak is of order $\Delta k \approx 10^{-26}$ and the most probable curve length is essentially the classical value 
$L=N\epsilon$. 

In general we may conclude that a macroscopic ``cold" curve has a well defined length of about the classical value. Only for much shorter curves and/or ``hot" curves does the quantum nature becomes relevant; shorter means only a few line elements and cold means a temperature very roughly a hundred times less than the Planck energy. 

\subsection{Summary comments on curve thermodynamics}

In this section we have pursued a statistical thermodynamic treatment of a curve. Obviously this section is more speculative than the previous sections since it requires introducing an effective curve energy and temperature. Our choice of the energy, taken to be proportional to the length of the curve, is reasonable but certainly not unique; for example we could choose an energy proportional to the square of the curve length and would obtain similar results. 

We were led to infinite and negative temperatures by standard statistical thermodynamic arguments. Such concepts are not unique to our theory, but also are encountered in spin systems and pumped lasers, for example. The infinite temperature needed to observe zero or negative length curves partly resolves the question of what a negative length means physically, but basic questions remain. For example how does one ``heat up" a curve or measure its length, and should we expect it to be in equilibrium with its environment, etc. We will return to these questions in the summary section. 

A more specific question concerns the fundamental energy $E_f$. On purely dimensional grounds it is reasonable to use the Planck energy $E_{pl}$ and of course take the length parameter $\epsilon$ to be the Planck distance $l_{pl}$. But it is also plausible that $E_f$ differs greatly from the Planck energy. This is because there are two length scales in spacetime physics, the Planck scale $l_{pl} \approx 10^{-35}m$ and the deSitter or Hubble scale $L_{H} \approx 10^{26}m$, whose ratio is $\approx 10^{-61}$, which has been called a gravitational fine structure constant or $\alpha_{g}$. \cite{alphag} Thus we might take $E_f \approx( {\alpha_g})^n E_{pl}$ with an arbitrary power $n$. In any case we certainly do not now see obvious evidence of a large Planck scale energy associated with a curve in spacetime; of course the vacuum energy associated with QFT poses similar unsolved problems. 

\section{Generally covariant theory}

It is easy to generalize the above ideas so they apply in covariant form to a general space, so long as it admits a Clifford algebra and thus a linear metric; the price is that we need an n-trad of vectors. We will illustrate this for four-dimensional spacetime. Thus we begin with a general line element, although we assume a signature $(+,-,-,-)$, 
\begin{equation}
\label{50}
ds_{cl}^2=g_{\mu \nu}dx^{\mu} dx^{\nu} .
\end{equation}
In the spacetime we introduce a tetrad of vectors $e_a ^ \mu $ labeled by indices $a$ etc. near the beginning of the alphabet and normalized in terms of the Lorentz metric as follows, 
\begin{equation}
\label{51}
e_a ^ \mu g_{\mu \nu } e_b ^ \nu = \eta _{ab}
\end{equation}
Eq. (\ref{51}) can be solved for the metric $g_{\mu \nu}$  in terms of the Lorentz metric $\eta_{ab}$  if we denote the inverse of the matrix $e_a ^ \mu$ as $\bar e_\mu ^ b$. That is 
\begin{equation}
\label{52}
\bar e_\mu ^ b e_b ^ \nu=\delta _\mu ^ \nu .
\end{equation}
Then we obtain
\begin{equation}
\label{53}
g_{\mu \nu}=\bar e_\mu ^ a \eta _{ab} \bar e_\nu ^ b .
\end{equation}
The matrix of tetrads $\bar e ^a _\mu$ thus serves as a sort of square root of the metric, somewhat analogous to the square root operation with the Dirac algebra. In terms of the tetrad the line element is thus
\begin{equation}
\label{54}
ds_{cl}^2=\eta_{ab} (\bar e^a_\mu\ dx^\mu)(\bar e_\nu^b dx^\nu)=\eta_{ab} d\tilde x ^a d\tilde x ^b, \; \; d\tilde x^a =\bar e^a_\mu dx^\mu ,
\end{equation}
so that in terms of the tetrad components $d \tilde x ^a$ the line element is explicitly
\begin{equation}
\label{55}
ds_{cl}^2=(d\tilde x^0)^2-(d\tilde x ^1)^2- (d\tilde x ^2)^2-(d\tilde x ^3)^2.
\end{equation}
From (\ref{55}) it is clear that the tetrad components are the geometrical proper times and ``meter stick distances" in a locally Lorentz system. We now define the linear line element as in sec.2 using the tetrad components $d\tilde x^a $ and the Dirac matrices from flat spacetime with Lorentz coordinates (\ref{6}), which we now denote by $\tilde \gamma _a$; that is
\begin{equation}
\label{56}
ds=\tilde \gamma_a d\tilde x ^a=\tilde \gamma_a (\bar e^a_\mu d x ^\mu)=(\tilde \gamma_a \bar e^a_\mu)d x ^\mu
=\gamma_\mu dx^\mu .
\end{equation}
This defines the covariant Dirac matrices $\gamma_\mu=(\tilde \gamma_a \bar e^a_\mu)$ in terms of the flat spacetime Dirac matrices; it then easily follows that the covariant Dirac matrices obey the Clifford algebra property
\begin{equation}
\label{57}
\{\gamma_\mu, \gamma_\nu \}/2=g_{\mu \nu} I .
\end{equation}
From (\ref{56}) and (\ref{57}) it follows as in sec.2 that the square of the linear metric is the metric times the identity matrix 
\begin{equation}
\label{58}
ds^2=\gamma_\mu \gamma_\nu dx^\mu dx^\nu=(1/2)\{ \gamma_\mu, \gamma_\nu \} dx^\mu dx^\nu = 
g_{\mu \nu}dx^\mu dx^\nu I ,
\end{equation}
and also
\begin{equation}
\label{59}
ds^2=\eta_{ab}d\tilde x^a d\tilde x^b I .
\end{equation}
Thus it is clear that our algebraic discussions in the previous sections should be applied to the line element expressed in terms of the tetrad components of the coordinate differentials $d\tilde x^a$. That is, when we take the line elements to be quantized in terms of some fundamental parameter it is obviously the tetrad components that we must use, so that $d\tilde x^a=n^a \epsilon$. 

To illustrate the covariant theory let us use the Schwarzschild metric. We align the tetrad with the coordinate directions $t, r, \theta, \phi$. Then the metric tensor and the triad matrix  $\bar e_{\mu}^a$ may be written in terms of the metric function $f(r)=1-2m/r$ as
\begin{equation}
\label{60}
g_{\mu \nu}= \left( \begin{array}{cccc} f & 0 & 0 & 0 \\ 0 & -1/f & 0 & 0 \\ 0 & 0 & -r^2 & 0 \\ 0 & 0 & 0 & -r^2 \sin^2 \theta \end{array} \right ),  \;
\bar e_{\mu}^a= \left( \begin{array}{cccc} \sqrt f & 0 & 0 & 0 \\ 0 & 1/ \sqrt f & 0 & 0 \\ 0 & 0 & r & 0 \\ 0 & 0 & 0 & r \sin\theta \end{array} \right ).  
\end{equation}
The tetrad components of the coordinate differentials are
\begin{equation}
\label{61}
d\tilde x ^0 = \sqrt f dt =n^0 \epsilon, \; d\tilde x^1= dr(1/\sqrt f )=n^1 \epsilon, \; d\tilde x^2 = rd\theta =n^2\epsilon, \; d\tilde x^3 = r \sin\theta d\phi =n^3 \epsilon . 
\end{equation}
Similarly the covariant Dirac matrices are 
\begin{equation}
\label{62}
\gamma_0 = \sqrt f \tilde \gamma_0 , \; \gamma_1= \tilde \gamma_1(1/\sqrt f) , \; \gamma_2 = r \tilde \gamma_2 , 
\;  \gamma_3 = r \sin\theta \tilde \gamma_3 . 
\end{equation}
Continuing in this way we obtain the linear line element in the form
\begin{equation}
\label{63}
ds = \gamma_\mu dx^\mu=\ \left( \begin{array}{cc} 0 & \sqrt f dt I + ds_3\\ \sqrt f dt I -ds_3 & 0  \end{array} \right), 
\end{equation}
where $ds_3$ denotes the three-space line element 
\begin{equation}
\label{64}
ds_3 = \left( \begin{array}{cc} r \sin \theta d\phi & dr/\sqrt f- ird\theta\\ dr/\sqrt f+ ird\theta & -r \sin \theta d\phi  \end{array} \right).  
\end{equation}
Alternatively we may write the linear line element in terms of the tetrad components of the displacements $d \tilde x^\mu =n^\mu \epsilon$ as
\begin{equation}
\label{65}
ds =\tilde \gamma_\mu n^\mu \epsilon =\ \left( \begin{array}{cc} 0 & n^\mu \sigma_\mu\\ n^\mu \tilde \sigma _\mu& 0  \end{array} \right)\epsilon , 
\end{equation}
where the sigmas are the same as in flat space. This line element of course has the same form (\ref{13}) as in flat spacetime. 

It is clear from this example that it is straight-forward to work out the covariant formulation of the theory. 

\section{Summary, conclusions and further speculations}

We began this work with a derivation of a linear line element as the square root of the quadratic line element of Riemannian geometry in both three and four dimensional flat spaces. The linear line element lead naturally to a quantum of distance and positive or negative distance eigenvalues. For a curve with many elements the idea of an entropy naturally presented itself, and the entropy in turn required a curve to have an energy and temperature. Standard statistical  thermodynamics then gave an elegantly simple relation between the temperature of a curve and its most probable length. In particular we found that curves have a Òheat-shrinkingÓ property - that is they shrink in length as the temperature is raised, with the shrinking becoming significant at a fundamental temperature scale that we tentatively identified as the Planck energy. At low temperature the entire theory reduces to classical Riemannian geometry; that is quantum effects vanish in the limit of zero temperatures. 

The theory is obviously speculative, but the use of a linear line element appears to be quite plausible and well founded, and the assignment of an entropy and temperature to a curve is no more strange than making geometry the agent of gravity, as is done in GR, or assigning an entropy and temperature to a black hole.   

So far the theory deals only with distances in empty space, with no particles or other bodies present. No dynamical equations are as yet used so the theory is presently only kinematical. However the heat shrinking property may perhaps be a vague hint that the theory might be used in a cosmogenesis scenario based on temperature: the Universe might start with an infinite or very high spacetime temperature, with all time and space intervals equal to zero, then cool down so that time begins. That is proper time stops as we go backwards toward a very high temperature that defines the ``beginning." 

Since our theory assigns an entropy to a curve considered as a string of spins it obviously has some similarity with thermogravity, in which entropy is assigned to two-dimensional surfaces. Thus it is possible that our analysis might be related to and perhaps justify some of the work on treating gravity as an emergent entropic force. \cite{Verlinde, Jacobson, emerge, Pady} 

Obviously much remains to be done to produce a complete theory: equations giving the line element operator for gravity, a discussion of areas and volumes, inclusion of objects in spacetime, a more complete consideration of observables of the theory, etc. So, unlike string theory that purports to be a theory of everything, our theory is one of empty spacetime without even a gravitational field. 

\section{Acknowledgements}

The seeds of this work date to the early1960s during discussions with the late Heinz Pagels when we wrote the equation for the linear line element of spacetime on the board and wondered what it meant, if anything.\cite{Pagels} More recently, colleagues in the Gravity Probe B theory group have provided patient criticism and stimulation, in particular Paul Worden, Robert Wagoner, and Francis Everitt. James Bjorken and I have had many interesting discussions and arguments on the general question of the fundamental nature of spacetime and LQG type theories, and Daniel Brandt has given me the interesting perspective of an experimentalist. Finally, Maarten Golterman detected several errors and provided interesting comments on the thermodynamical oddities of the theory. 
\appendix

\section{The Pauli equation and the electron magnetic moment}

We include appendices A and B as motivation for taking the square root of various operators. In non-relativistic QM the time independent Schr\"{o}dinger equation for a free particle, which is the same as the energy eigenvalue equation, is 
\begin{equation}
\label{66}
({\vec p}\ ^2 /2m)\psi =E \psi , \; p^j = -i\hbar \partial _j ,
\end{equation}
We take the square root of (\ref{66}) by using the Pauli algebra (\ref{3}) and find
\begin{equation}
\label{67}
(p^i \sigma_i)^2=(\vec p \cdot \sigma) ^2 = \vec p\ ^2 I .
\end{equation}
Then we replace the Schr\"{o}dinger equation (\ref{66}) for the single component wave function by promoting $\psi$  to be a two-component spinor wave function, and obtain
\begin{equation}
\label{68}
(1/2m)(\vec p \cdot \sigma) ^2 \psi= E \psi .
\end{equation}
Of course (\ref{68}) contains nothing new since it is the same as (\ref{66}) written twice; but we now couple the electron to the electromagnetic field $A^ \mu$  via the canonical minimal coupling recipe
\begin{equation}
\label{69}
E \rightarrow E-eA^0 , \; \vec p \rightarrow \vec p -e\vec A, 
\end{equation}
and obtain 
\begin{equation}
\label{70}
(1/2m)[(\vec p -e\vec A) \cdot \sigma] ^2 \psi= (E-eA^0)I \psi .
\end{equation}
The Pauli algebra in (\ref{3}) lets us put (\ref{70}) in the form
\begin{equation}
\label{71}
(1/2m)[(\vec p -e\vec A)^2 - e\hbar \vec B \cdot \sigma]  \psi= (E-eA^0) \psi .
\end{equation}
Eq.(\ref{71}) is the Pauli equation describing an electron in an electromagnetic field, with the correct g factor of 2, as is clear from  the magnetic interaction term. The g factor of 2 is often incorrectly considered  to be an effect of relativity, but is clearly the result of taking a square root using the Pauli algebra; the relativistic Dirac equation is not needed. 

\section{Dirac equation as a square root}

This appendix is a four-dimensional analog of the previous appendix A. We begin with the Klein-Gordon equation for a free particle, 
\begin{equation}
\label{72}
p^2\psi = m^2  \psi ,\; p_ \mu =i\hbar \partial_\mu .
\end{equation}
Then using the Dirac algebra we observe that 
\begin{equation}
\label{73}
(p^\mu \gamma_\mu)^2=p^2 I,
\end{equation}
so the Klein-Gordon equation implies
\begin{equation}
\label{74}
(p^\mu \gamma_\mu)^2\psi=m^2 I \psi.
\end{equation}
Thus we may promote $\psi$ to be a four component spinor and take the square root of (\ref{74}) to obtain
\begin{equation}
\label{75}
(p^\mu \gamma_\mu  \pm m) \psi = 0. 
\end{equation}
Finally we couple the system to the electromagnetic four-vector potential $A^\mu$ with the standard minimal coupling recipe in four dimensions (\ref{69}) and obtain,
\begin{equation}
\label{76}
[(p^\mu-eA^\mu) \gamma_\mu  - m I] \psi = 0, 
\end{equation}
the Dirac equation for an electron in an electromagnetic field.
 
\section{Line element eigenvectors}

The eigenvalues of the three-dimensional line element operator (\ref{7}) are $\pm \epsilon$ and the eigenvectors may be easily obtained in the standard way; one convenient way to express them, easily verified, is in the normalized form,
\begin{equation}
\label{77}
|+, \hat n \rangle= \left( \begin{array}{c} \cos(\theta/2)e^{-i\phi /2} \\ \sin(\theta/2)e^{i\phi/2} \end{array} \right), \;
|-, \hat n \rangle= \left( \begin{array}{c} -\sin(\theta/2)e^{-i\phi /2} \\ \cos(\theta/2)e^{i\phi/2} \end{array} \right). 
\end{equation} 
This displays the half angle nature of spin rather elegantly, and the notation will be convenient in obtaining the eigenvectors of the line element operator in four-dimensional spacetime. 

For four-dimensional spacetime we will carry out the eigenvalue solution in detail for the case $\lambda =1$ and $\eta =1$, which corresponds to the path of a massive particle moving at less than $c$. Denoting the eigenvector by $V$ we have
\begin{equation}
\label{78}
(n^\mu \gamma_{\mu} )V=\left( \begin{array}{cc} 0 &  n^{\mu}\sigma_{\mu} \\ n^{\mu} \tilde\sigma_{\mu} & 0 \end{array} \right) \left(\begin{array}{c} \alpha \\ \beta \end{array} \right ) 
=\left(\begin{array}{c} \alpha \\ \beta \end{array} \right ), 
\end{equation}
where $\alpha$ and $\beta$ are the 2-tuple parts of $V$. From (\ref{78}) we thus have
\begin{equation}
\label{79}
\beta=n^{\mu}\tilde \sigma_{\mu}\alpha=\large (n^0 -n (\vec n \cdot \sigma) \large )\alpha, \; n=|\vec n|.
\end{equation}
The degeneracy of the eigenvalues $\lambda$ allows a choice of the 2-tuple $\alpha$; we take for convenience the natural choice $|+, \hat n\rangle$ from (\ref{77}) and obtain
\begin{equation}
\label{80}
\beta = (n^0 - n)|+, \hat n\rangle
\end{equation}
and for the eigenvector 
\begin{equation}
\label{81}
V= \left(\begin{array}{c} |+, \hat n\rangle \\ (n^0 - n)|+, \hat n\rangle \end{array} \right )
=|+1, +, \vec n \rangle .
\end{equation}
The form (\ref{81}), not normalized, is useful for our present purposes since we are interested in the special cases $n=0$ and $n^0=0$. (We could also use hyperspherical angles, which would give a bit more elegance.) In (\ref{81}) we have labelled the state as  $|\lambda, \pm, \vec n \rangle $ with $\lambda =+1$ for the important eigenvalue, and + and $\vec n$ to label the $\alpha$ state. It is useful that the eigenvector in four-dimensional spacetime may be written so simply in terms of the eigenvalue $\lambda$ and the eigenvector of the three-dimensional problem. 

The other eigenvectors are obtained in a similar way. For the timelike case they are
\begin{equation}
\label{82}
|+1, \pm, \hat n \rangle=\left(\begin{array}{c} |\pm, \hat n\rangle \\ (n^0 \mp n)|\pm, \hat n\rangle \end{array} \right ), \;
|-1, \pm, \hat n \rangle=\left(\begin{array}{c} |\pm, \hat n\rangle \\ -(n^0\mp n)|\pm, \hat n\rangle \end{array} \right ),
\end{equation}
and for the spacelike case, 
\begin{equation}
\label{83}
|+i, \pm, \hat n \rangle=\left(\begin{array}{c} |\pm, \hat n\rangle \\ -i(n^0\mp n)|\pm, \hat n\rangle \end{array} \right ), \;
|-i, \pm, \hat n \rangle=\left(\begin{array}{c} |\pm, \hat n\rangle \\ i(n^0\mp n)|\pm, \hat n\rangle \end{array} \right ).
\end{equation}

For the null case, instead of (\ref{78}) and (\ref{79}) the eigenvalue equation gives  
\begin{equation}
\label{84}
\large (n^0 -n (\vec n \cdot \sigma) \large )\alpha = 0 , \; \large (n^0 +n (\vec n \cdot \sigma) \large )\beta = 0.
\end{equation}
Since $n^{\mu}$ is a null vector there are two possibilities to consider, $n=n^0$, which we label as $\lambda =+0$,
and $n=-n^0$, which we label as $\lambda =-0$ in analogy with the previous cases. For $\lambda = +0$ (\ref{84}) becomes 
\begin{equation}
\label{85}
\large (I -(\vec n \cdot \sigma) \large )\alpha = 0 , \; \large (I + (\vec n \cdot \sigma) \large )\beta = 0,
\end{equation}
and we see by inspection that there are two solutions, 
\begin{equation}
\label{86}
 |+0, +, \hat n \rangle=\left(\begin{array}{c} |+, \hat n\rangle \\ 0 \end{array} \right ), \; 
 |+0, -, \hat n \rangle=\left(\begin{array}{c} 0 \\ |-, \hat n \rangle \end{array} \right )
 \end{equation}
Similarly for $\lambda = -0$ the two solutions are
\begin{equation}
\label{87}
 |-0, +, \vec n \rangle=\left(\begin{array}{c} 0 \\ |+, \hat n \rangle \end{array} \right ), \; 
 |+0, -, \vec n \rangle=\left(\begin{array}{c} |-, \hat n \rangle \\  0\end{array} \right ).
 \end{equation}
Because of the degeneracy the above solutions are of course not unique and not always the most convenient.

\begin{figure}[htbp] 
  \centering
   \includegraphics[width=4in]{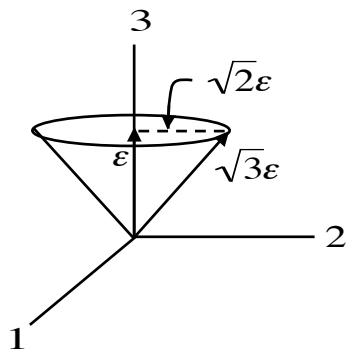} 
   \caption{Transverse uncertainty in the line element due to its operator nature.}
   \label{fig1}
\end{figure}
\begin{figure}[htpb] 
     \centering
     \includegraphics[width=5in]{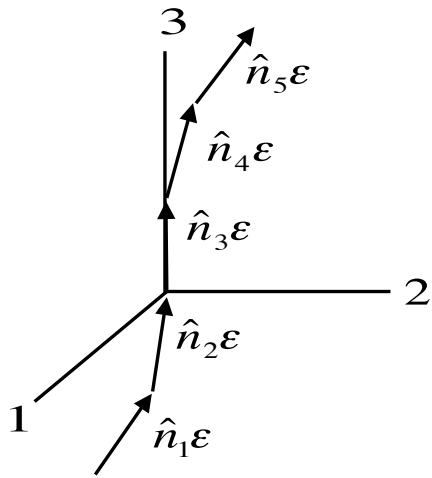} 
   \caption{Curve considered as a direct sum or chain of independent spins}
   \label{fig2}
\end{figure}
\begin{figure}[htbp] 
     \centering
     \includegraphics[width=5in]{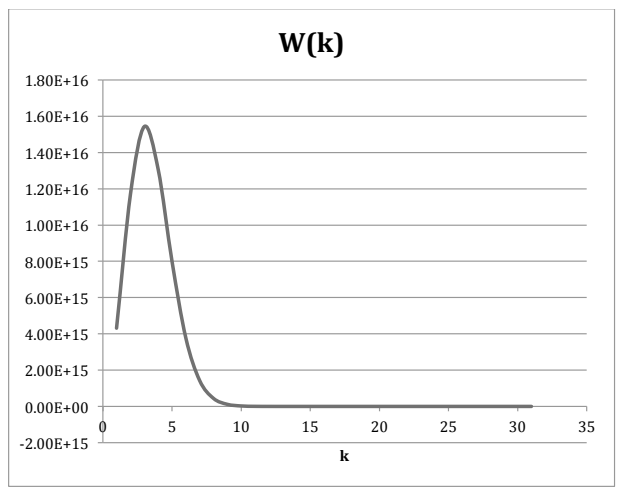} 
   \caption{Showing the sharp peak of $W(k)$ at the most probable value of $k$, illustrated for $N=30$ 
   and  $E_f /T = 1.4$,}
   \label{fig3}
\end{figure}

\end{document}